\documentclass[preprint,showpacs,aps,pre]{revtex4}
\usepackage{amsmath}
\usepackage{amssymb}
\usepackage{graphicx}
\usepackage{graphics}

\begin{document}

\title{Kinetic mechanism of chain folding in polymer crystallization}
\author{S. Stepanow}
\affiliation{Martin-Luther-Universit\"{a}t Halle-Wittenberg, Institut f\"{u}r Physik,
D-06099 Halle, Germany }
\date{\today}

\begin{abstract}
I develop a kinetic mechanism to explain chain folding in polymer
crystallization which is based on the competition between the formation of
stems, which is due to frequent occupations of trans states along the chains
in the supercooled polymer melt, and the random coil structure of the
polymer chains. Setting equal the average formation time of stems of length $%
d_l$ with the Rouse time of a piece of polymer of the same arc length $d_l$
yields a lower bound for the thickness of stems and bundles. The estimated
lamellar thickness is inversely proportional to the supercooling. The
present approach emphasizes the importance of repulsive interactions in
polymer crystallization, which are expected to be responsible for the
logarithmic lamellar thickening and the increase of lamellar thickness with
pressure. An expression for the growth rate for formation and deposition of stems is derived by considering the
growth as a dynamic multistage process.
\end{abstract}

\maketitle



\section{Introduction}

The understanding of polymer crystallization and its theoretical description
in the framework of polymer statistics remains a challenging problem since
its discovery \cite{keller57}-\cite{till57}. The classical theories of
polymer crystallization \cite{lauritzen-hoffman60}-\cite{hoffman76} (see
also \cite{armitstead92}-\cite{mandelkern-Buch} and citations therein) are
based on the nucleation theories developed for low molecular weight systems.
The lamellar thickness is identified ad hoc with the critical size of
nuclei, and is not related with the key features of polymers in melts and
solutions. Despite enormous interest in polymer crystallization over many
decades, the molecular mechanism of polymer crystallization is not
understood, and a polymer statistics related description is not available. A
review of the research in the field of polymer crystallization in the
eighties can be found in \cite{fdisc79}. Many important specific features of
polymer crystallization were established in the recent research in the
field, which includes experimental work \cite{imai93}-\cite{panine08},
computer simulations \cite{kavassalis95}-\cite{luo-sommer11}, and theoretical
studies \cite{dimarzio-passaglia87}-\cite{kundagrami-muthu07} (and citations
therein). The progress in the field of polymer crystallization in recent
years is reviewed in \cite{sommer-reiter03}-\cite{piorkowska-rutledge13}.

A new view of polymer crystallization was developed by Strobl \cite{strobl00}-\cite{strobl-rmp09}, who started from the observation that the nucleation
based theories are in disagreement with experiments \cite{strobl98}-\cite%
{rastogi91}. The dependence of the crystallization temperature $T$ on the
inverse lamella thickness $d^{-1}_l$ has a larger slope than that of the
melting temperature, so that these curves intersect at some temperature
(comparable with the temperature of zero growth $T_{zg}$), which implies
that the polymer crystallization can develop only below $T_{zg}$. The
picture proposed by Strobl is based on a multistage character of the
crystallization process, and the existence of a mesomorphic layer as
precursor to the crystalline phase without however specifying the
statistical mechanical origin of his scenario.

The aim of this article is to develop a description of polymer
crystallization on a more microscopic level relying on polymer specific
properties, e.g., the coil structure of the polymers, and in terms of the
relevant microscopic interactions between the monomers in the supercooled
polymer melt. I develop a kinetic mechanism of chain folding in polymer
crystallization which is based on the competition between the formation of
stems and the random coil shape of polymer chains, and is based on the view
that the driving force is due to the repulsions between fluctuational stems,
which form below a characteristic temperature $T_m^0$ due to favored
occupation of the trans states along chains, and orient in order to minimize
the excluded volume.
The importance of repulsive
interactions in this approach is in accordance with the general evidence of
the role of repulsive interactions in liquid-solid phase transformations.
The dynamic interplay between forming stems and Rouse dynamics considered in
the present approach is expected to manifest itself as specific fluctuations
prior formation of polymer crystals, and might be responsible for the
mesomorphic layer postulated by Strobl. The relevance of density and
orientation fluctuations for polymer crystallization is not new and is the
start point of the spinodal decomposition theories for the description of
early stages of polymer crystallization in \cite{olmsted98}, \cite{doi88}.
To cite \cite{gee06} "These (\emph{density and orientation}) fluctuations
are caused by an increase in the average length of rigid trans segments
along the polymer backbone during the induction period". However, the
lamellae thickness in these approaches was not brought in connection with
polymer properties.

The article is organized as follows. Section \ref{lamella} introduces the
basic ideas of the kinetic mechanism of chain folding. Section \ref{gr}
introduces to the calculation of the growth rate. Section \ref{concl}
summarizes our conclusions.

\section{Kinetic mechanism of chain folding}

\label{lamella}

Fluctuational occupations of the trans states in an supercooled polymer melt
of interpenetrating chains result in formation of stems possessing a finite
lifetime. The repulsion between the neighbor stems forces them to orient
parallel to each other in order to minimize the excluded volume and results
in formation of bundles. The increase of average occupation of trans states
below a temperature $T_m^0$ increases the lifetime of stems, and enforces
the effect of repulsions. The orientations of stems due to their mutual
repulsions is similar to the mechanism of the isotropic-nematic transition
in lyotropic liquid crystals, where according to Onsager \cite{onsager49}
the minimization of the excluded volume is responsible for the transition.
The difference between liquid crystals and polymers is that in the case of
polymer crystallization the stems do not exist from the beginning, but
emerge due to occupation of trans states, and orient and grow due to
repulsive interactions between the stems. The present approach is in
accordance with simulations in \cite{meyer01}-\cite{meyer02}, where chain
folding takes place in polymer crystallization by taking into account only
repulsive intermolecular interactions. The intramolecular dihedral energies
are associated with the differences between, e.g., gauche and trans states,
and are responsible for the formation of stems.

The length scale of forming fluctuational stems, which are forced to orient
due to mutual repulsive interactions, is determined by the competition
between the growth of stems and their relaxation due to Rouse dynamics. For
the average formation time of stems we adopt the following phenomenological
expression
\begin{equation}  \label{fs1}
\tau_s=\frac{d_l}{v_0 +c\Delta T},
\end{equation}
where $\Delta T=T_m^0-T$ is the supercooling and $v_0$ and $c$ are
constants.
The term in (\ref{fs1}) which is proportional to $\Delta T$ means that nonzero supercooling is necessary for stem growth. The non-zero value of $v_0$ accounts for the effect of an
orienting crystal surface on the formation of stems and bundles, and is thus
legitimate for secondary crystallization and in heterogeneous nucleation at
small supercooling, where the crystal growth begins at seeds.
Eq.~(\ref{fs1}) yields for the ratio $G_s=d_l/\tau_s$ the expression $%
G_s=v_0+c\Delta T$ for the longitudinal growth rate of stems. A similar
expression for the lateral growth rate for small supercooling is well-known
in the literature (see for example \cite{kundagrami-muthu07}). Because
stems at the time scale $\tau_s $ are expected to form and decompose, the
quantity $G_s$, which has the dimensionality of velocity rate, is a
fluctuational quantity, so that the above estimate has to be understood as
the typical value of $G_s$.

\begin{figure}[tbph]
\begin{center}
\includegraphics[scale=0.30]{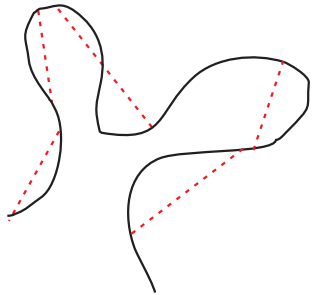}
\end{center}
\caption{(Color online) Visualization of fluctuating stems along the polymer coil. The two
first and two last stems are likely to constitute longer stems, while the
third stem (from the left) is likely to build a fold with the 2nd stem.}
\label{coil-and-stems}
\end{figure}

The Rouse time of a polymer with arc length $d_l$ is given by
\begin{equation}  \label{tRouse}
\tau_{Rouse}(d_l)=\frac{\zeta d_l^2}{3\pi^2k_BT_m^0},
\end{equation}
where $\zeta$ is the monomer friction coefficient ($\zeta \simeq 4.74 \times
10^{-13} Ns/m$ for polyethylene), and we have replaced $T$ in (\ref{tRouse})
by $T_m^0$ ($\simeq 135 ^{\circ}{C}$ for polyethylene), which is legitimate
for small supercooling. The quantity $k_B T_m^0/\zeta$ is the monomer
diffusion coefficient. The balance between the stem growth and their spatial
orientations, which is determined by the coil structure of the polymer, can
be expressed as follows
\begin{equation}  \label{ts}
\tau_s \simeq \tau_{Rouse}(d_l).
\end{equation}
Resolving (\ref{ts}) with respect to $d_l$ we obtain the characteristic
length scale determined by the interplay between the stem growth and the
random coil structure of polymer chains as
\begin{equation}  \label{dl-ineq}
d_l \simeq \frac{3\pi^2k_BT_m^0}{\zeta c(T_c^0-T)},
\end{equation}
where $T_c^0=T_m^0+v_0/c$. Eq.~(\ref{dl-ineq}) gives an estimate of the lower bound of the lamellar thickness.
The experimental lamellar thickness is always larger than that given by Eq.~(\ref%
{dl-ineq}), because two subsequent stems along a polymer can fold or form
one stem with some probabilities (see Fig.~\ref{coil-and-stems}).
Other processes e.g. those responsible for lamellar thickening also result in increase of $d_l$.
Resolving Eq.~(\ref{dl-ineq}) with respect to $T$ we arrive at the following relation
between the crystallization temperature and the lamellar thickness
\begin{equation}  \label{T-dl1}
T=T_c^0-\frac{3\pi^2k_B T_m^0}{c\zeta}\frac{1}{d_l}.
\end{equation}
The $T-d_l^{-1}$ relation without replacing $T$
in the expression of the Rouse time by $T_m^0$ reads $T=T_c^0/(1+3\pi^2k_B/%
\zeta c d_l^{-1})$.

Let us compare the orientation of stems in polymer crystallization with the
isotropic-nematic transition in lyotropic liquid crystals, where the
transition is determined by the condition $\Delta S_{or}+ \Delta S_{trans}=0$%
, where $\Delta S_{or}=k_B\ln(\Omega_n/\Omega_i)\simeq -k_B$ and $\Delta
S_{trans}=k_BnL^2D$ ($n$ is the density of rods, $L$ their length, and $D$
the transverse size) are the decrease of orientation entropy and the
increase of the translational entropy, respectively. Because the forming
stems in the polymer melt overlap, the above Onsager condition is fulfilled
and the stems can directly orient due to repulsive interactions. The van der
Waals interactions will be enhanced in bundles and contribute to their
stabilization.
The present kinetic mechanism of folding favors the
switchboard fold surface, which is the consequence of the coil structure of
polymers. The formation of stems and lamellae can be visually interpreted as
space segregation of the trans and gauche states. It is intuitively expected
that the repulsive forces promote a local disentanglement of interpenetrated
chains, while the attractive van der Waals interactions favor an amorphous
state, and are therefore expected to be only important for stabilization of
the bundles and consequently of the lamellar structure. The van der Waals
forces between polymer pieces outside the bundles and with the fold surface
would result in their adsorption on the latter. Therefore, we expect that
the (logarithmic) thickening of bundles and lamellae is also caused by
repulsive interactions. The enthalpic contributions to the processes of
polymer crystallization are due to i) the energy difference between trans -
and gauche states, and ii) the difference of the contribution of van der
Waals forces in the amorphous and the lamellar crystalline states.

The condition $T_c^0>T_m^0$ is a consequence of the assumption $v_0 \neq 0$
in the Ansatz in Eq.~(\ref{fs1}). The van der Waals attraction of formed
stems to the surface is expected also to contribute to the non-zero value of
$v_0$. The estimate of $\tau_{Rouse}$ for $d_l=15 \,$nm yields for example
for polyethylene the value of order of $10^{-9} \,s$. Setting $\tau_s $
given by Eq.~(\ref{fs1}) equal to $\tau_{Rouse}$ expresses the interplay
between the isotropic-nematic ordering of emerging stems and the coil
structure of polymers. The qualitative picture is that that
the stems grow until the parts of a chain outside a bundle, which are also
forced to form stems, will be located on the lateral side of the bundle, and
will likely belong to the same bundle (see Figure \ref{coil-and-stems}).
Thus, the competition between the growth and orientation of stems and the
coil structure of the polymer chains yields that the consecutive stems along
the polymer fold and belong to the same bundle. Therefore, the longitudinal
growth of the bundles is restricted by the coil structure of polymers, which
applies in the Rouse theory on all scales. The slower (logarithmic)
thickening of bundles can further occur. At larger supercooling, a smaller
stem length will be selected as a result of this interplay. The kinetic
mechanism of chain folding suggests that the lamellar thickness $d_l$
follows the change of supercooling. The latter is in accordance with
observations made long ago \cite{bassett-keller62}. The chains in polymer
melts are Gaussian coils irrespective of the presence of entanglements, so
that according to the above picture we expect that entanglements weakly
influence the condition (\ref{ts}). However, the influence of entanglements
on the growth rate is more complex (see \cite{mandelkern-Buch}, Vol II,
Sect. 9.14 and \cite{klein-ball79}). The proposed folding mechanism is
expected to apply for both primary and secondary crystallization processes.
For primary crystallization, where the surface effect is marginal ($v_0
\simeq 0$), one can apply (\ref{T-dl1}) with $T_c^0=T_m^0$. We expect that
the above folding mechanism applies for crystallization from polymer
solution too, where the slow collapse due to van der Waals interactions
occurs first, which is followed by the fast folding mechanism due to
repulsion-orientation coupling after the repulsive interactions become
significant as a result of increase of density. We also expect that the
above folding mechanism based on trans and gauche conformation states is
generic for polymers with more complicated local conformation states.

Eq.~(\ref{ts}) implies that the folding length $d_l$ is selected in the fluctuational regime associated with the
(`microscopic') time $\tau_s$. Because the Ansatz given by Eq.~(\ref{ts}) is
local, the lamellar thickness is expected, in contrast to the growth rate,
to be robust with respect to changes of external parameters such as
molecular weight, etc. This consequence of the Ansatz in Eq.~(\ref{ts}) is
in accordance with experiments \cite{mandelkern-Buch}. The lamella thickness
$d_l$ lies for melts or solutions in the range between $10-20 ~$nm, and
shows a weak dependence on the moderate increase of pressure \cite%
{wunderlich63}, whereas the melting temperature significantly increases
under pressure.

The influence of external pressure on polymer crystallization shows \cite%
{wunderlich64}, \cite{mandelkern-Buch}, Vol.II, Fig. 12.8 (spherulites) that
the melting temperature increases with pressure. The lamellar thickness
increases smoothly for moderate pressures \cite{wunderlich64}. At large
pressures the lamellar thickness increases considerably and can achieve a
few microns, and approaches that of an extended crystal \cite{hikosaka05}. A hexagonal phase
was observed for large pressure \cite{ungar-keller80}, which has at the coexistence curve a lower density
than the orthorhombic phase \cite{bassett81}, p. 171. These observations
emphasize the importance of the repulsive interactions in polymer
crystallization. This is similar to the van der Waals gas where the increase
of the pressure shifts the interplay between the repulsive and attractive
interactions in favor of the former. The responsibility of the repulsive
interactions for stability of the hexagonal phase in polymer crystals is
similar to the formation of the triangular lattice of flux lines in type II
superconductors, which has its origin in repulsive interactions between the
flux lines \cite{nelson89}. The repulsive interactions in the spatially
ordered stems in the lamellae facilitates the sliding diffusion, which is
expected to be responsible for the increase of lamellae thickness at large
pressures \cite{hikosaka05}. Thus, the increase of the lamellar thickness,
the development of the hexagonal phase at high pressures are direct
evidences of the importance of repulsive interactions for stem formation in
polymer crystallization.

Since $T_c^0 >T_m^0$ and because of fact that crystallization can occur for $%
T_c(d_l)<T_m(d_l)$, the crystallization line, which is described by Eq.~(\ref%
{T-dl1}), has to cross the melting line, which is given by the Gibbs-Thomson
relation
\begin{equation}  \label{gt}
T =T_m^0 -\frac{2\sigma_eT_m^0}{\Delta h}\frac{1}{d_l},
\end{equation}
where $\Delta h$ is the heat of fusion and $\sigma_e$ is the surface tension
of the fold surface.

\begin{figure}[tbph]
\begin{center}
\includegraphics[scale=0.40]{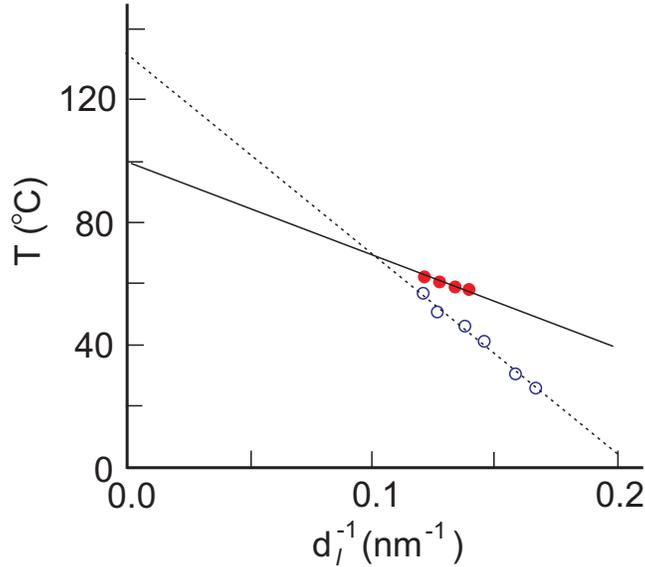}
\end{center}
\caption{(Color online) Open circles: experimental data for poly($\protect\epsilon$%
-caprolactone) from \protect\cite{heck-strobl-njp99}. Dashes: Our fit of the
crystallization line. Filled circles and the continuous line: the
melting line.}
\label{T-dl-fit}
\end{figure}

Note that Eq.~(\ref{T-dl1}) with $T_c^0>T_m^0$ is in accordance with
Strobl's analysis of experimental data.
The parameters in Eq.~(\ref{T-dl1}) can be estimated from the fit to the crystallization line
using the experimental data from \cite{heck-strobl-njp99}, which is shown in
Figure \ref{T-dl-fit}.
The ratio $v_0/c$ is equal to $T_c^0-T_m^0$ and possesses for poly($\epsilon$%
-caprolactone) according to Fig.~\ref{T-dl-fit} the value $36$ K. The slope
of (\ref{T-dl1}), which is obtained as $\partial T/\partial
d_l^{-1}=-3\pi^2k_BT_m^0/c\zeta$, possesses according to \ref{T-dl-fit} the
value $-650$ K nm. Thus, Eq.~(\ref{T-dl1}) with constants $v_0\zeta$ and $%
c\zeta$ estimated for poly($\epsilon$-caprolactone) as $v_0\zeta \simeq 610
\,k_B$ and $c\zeta \simeq 17 \,k_B$ coincides with the crystallization line
in Fig. \ref{T-dl-fit} (Fig.~10 in \cite{strobl-rmp09}, and Fig. 11 in \cite%
{heck-strobl-njp99}).
It is likely that uncertainties of the experimental data are responsible for not too excellent fit in Fig. \ref{T-dl-fit}.

\section{Growth rate}

\label{gr}

We now will consider the time evolution of formation and deposition of stems
in the vicinity of the crystallization front. A self-consistent treatment of
the mutual correlations of different stems enables one to consider a
time-dependent growth rate of one stem $G(t)$. At the time $t=t_l$, when, on
average, one stem attaches to the crystal surface, the quantity $G \equiv
G(t=t_l)$ is the growth velocity given by the ratio of one attached stem per
time $t_l$. The multistage character of the growth means that the growth
occurs by sequences of processes, and implies that $G(t+\Delta t)$ depends
on $G(t)$, which enables one to write down the following phenomenological
differential master-type equation for $G(t)$ as function of time
\begin{equation}  \label{g1}
\frac{dG}{dt}=-\gamma G t^{-\alpha},
\end{equation}
where $\alpha<1$ ($\alpha=1/2$ in the following) and $\gamma$ is a constant.
The factor $t^{-\alpha}$ in (\ref{g1}) takes into account the slow-down of
the variation of $G$ with time, which is expected to be due to the random
character of the attachment process.
The absence of a positive term on the right-hand side of (\ref{g1}) ensures
that Eq.~(\ref{g1}) does not possess a steady state solution. Note that
master-type equations for crystal growth (see e.g. \cite{sadler86}) describe
many stems growth, and do possess a steady state solution. Eq.~(\ref{g1}) is
similar to the multiplicative renormalization of quantities from microscopic
to macroscopic scales in the theory of critical phenomena in the case when
the coupling constant does not renormalize \cite{zinn-justin}. Because Eq.~(%
\ref{g1}) is intended to describe the growth rate of one stem it should be
integrated from $t=0$ until the time $t=t_l$, which corresponds to formation
and attachment time of one stem to the crystalline front. The integration of
Eq.~(\ref{g1}) from $t=0$ until the time $t=t_l$ yields
\begin{equation}  \label{g1a}
G \equiv G(t_l)=G_0\exp\left( -2\gamma \sqrt{t_l} \right).
\end{equation}
For comparison, the growth rate for ballistic deposition is independent of $%
t $, while $G$ for diffusion controlled deposition is proportional to $%
t^{-1/2} $. The above equation suggests that $t_l$ is infinite at the onset
of crystallization, where the growth rate is zero i.e. $G(t_l \simeq
\tau_s)=0$ at $T=T_c^0$. 
Note that in contrast to (\ref{g1a}) with $t_l \simeq \tau_s$ the expression
for the longitudinal growth rate, which follows from Eq.~(\ref{fs1}), is
given by $G_s=d_l/\tau_s=v_0+c\Delta T$.

A naive identification of $t_l$ with the time given by the condition (\ref%
{ts}) yields
\begin{equation}  \label{ts1}
t_l \simeq \tau_s=\frac{d_l}{c(T_c^0-T)}=\frac{3\pi^2k_BT_m^0}{c^2\zeta}%
(T_c^0-T)^{-2}.
\end{equation}
However, as consequence of the intersection of the crystallization and
melting lines given by Eqs.~(\ref{T-dl1}-\ref{gt}) shown in Fig. \ref%
{T-dl-fit} the growth rate is non-zero only below the intersection
temperature $T_{is}$. The latter is also in accordance with Strobl's
analysis of experimental data \cite{strobl-rmp09}, where the growth rate
becomes zero at the temperature $T_{zg}<T_m^0$. Thus, to take this
circumstance into account we adopt (\ref{ts1}) with $T_c^0$ replaced by $%
T_{zg}$ which yields the time $t_{g}$, which is larger than $t_l$. The
difference between $t_l$ and $t_{g}$ can be understood as follows: While $%
t_l $ ($\simeq\tau_s$) gives the selection rule for lamella thickness from
the comparison of time scales of competing processes (i) stem formation and
(ii) coil shape of polymers, and does not make a statement on the time
course of the growth process, the time $t_{g}$ is associated with the real
time of formation and attachment of a stem at the crystal surface by taking
into account the complicated dynamics, and is therefore much larger than $t_l
$. Note that the orientation time of stems is not included in $t_l$, but in $%
t_g $. Inserting $t_g$ for $t_l$ in (\ref{g1a}) we arrive at the following
estimate of the growth rate
\begin{equation}  \label{g3}
G =\tilde{G}_0 \exp\left(-\frac{a}{T} -2\gamma\sqrt{\frac{3\pi^2k_BT_m^0}{%
c^2\zeta}}\frac{1}{T_{zg}- T}\right),
\end{equation}
where $G_0=\tilde{G}_0\exp(-a/T)$ is introduced to take into account the
increase of the relaxation time (viscosity) 
with decrease of temperature. It follows from Eq.~(\ref{g3}) that $G$
possesses a pronounced maximum as a function of $T \leq T_{zg}$. Note that
in contrast to the Turnbull-Fisher expression \cite{turnbull-fisher49} for $%
G $ in nucleation theory, the above expression describes the attachment rate
of one stem. To obtain the experimentally measured growth rate one should
multiply (\ref{g3}) with the average number of stems formed per time and per
volume. The quantity $(2\gamma/\sqrt{c})\sqrt{3\pi^2k_BT_m^0/c\zeta}%
=51\gamma/\sqrt{c}$ is equal to the characteristic temperature $T_G$
appearing in the growth rate $u = u_0 \exp \left(-T^{*}_A/T-T_G/(T_{zg}-T)
\right)$ given by Eq.~(6) in \cite{cho-strobl07}. The fit for poly($\epsilon$%
-caprolactone) yields $T_G =397 \, ^{\circ }C$, so that one obtains $\gamma/%
\sqrt{c}\simeq 7.8$.

Note that the existence of two separated time scales $\tau_s$ and $t_g$,
which have a clear physical meaning in the present approach, is in
accordance with the experimental finding that the growth rate depends
exponentially on $t_g$ and $\gamma$, which is expected to depend
considerably on external parameters such as pressure, molecular weight,
entanglements, etc. \cite{mandelkern-Buch}, while the lamellar thickness is
determined by the `microscopic' time $\tau_s$, which, as it follows from the
definition, is less sensitive to the external parameters.

\section{Conclusions}

\label{concl}

To summarize, I developed a kinetic mechanism for polymer folding in polymer
crystallization, which is based on the competition between the stem
formation, which is the consequence of preferential occupations of trans
states in the supercooled polymer melt, their orientation to minimize the
excluded volume, and the coil shape of polymer chains. In contrast to the
phenomenological nucleation based theories, where the size of critical
nucleus is identified ad hoc with the lamellar thickness, our approach has
the aim to understand the basic features of polymer crystallization in terms
of microscopic interactions between the monomers in supercooled polymer
melts, and the coil structure of polymers. The present approach suggests
that the selection of lamella thickness in polymer crystallization is of
kinetic origin, and is determined by the `microscopic' time scale $\tau_s$.
The (logarithmic) lamellar thickening as well as the increase of the
lamellar thickness with pressure substantiate the importance of the
repulsive interactions in polymer crystallization. The dynamic interplay
between forming stems and Rouse dynamics considered in the present approach
is expected to manifest itself as an ordered precursor prior formation of
polymer crystals, and might be responsible for the mesomorphic layer
postulated by Strobl.

Note that the present kinetic mechanism of chain folding is in accordance with results of very recent molecular dynamics simulation \cite{schilling2014} on the time sequence of basic processes in formation of the crystalline order: "first the chain segments align, then they straighten, and finally the cluster become denser and local positional and orientational order are established".
Further, the finding in \cite{schilling2014} that entanglements do not affect the nucleation but the growth process is also in accordance with the main conclusion of the present work that there are well separated time scales responsible for selection of the lamellar thickness and formation and attachment of stems i.e. the crystal growth.

The growth rate for formation and deposition of one
stem at the growing surface is derived from the differential master-type
equation for the scale dependent attachment rate of one stem, which
incorporates the multistage character of the secondary crystallization. The
implementation of the picture of polymer folding in polymer crystallization
proposed in this article in coarse grained analytic and numeric models will
allow more quantitative predictions on polymer crystallization.

\begin{acknowledgments}
I acknowledge partial financial support from the German Research Foundation,
Ste 981/3-1 and SFB TRR 102. I am grateful to W. Paul, K. Saalw\"{a}chter,
C. Schick, and T. Thurn-Albrecht for useful discussions.
\end{acknowledgments}

\end{document}